\documentclass{emulateapj}
\usepackage{apjfonts}

\slugcomment{To appear in ApJ Letters}

\shorttitle{Spitzer IRAC imaging of PKS 0637$-$752 Jet}
\shortauthors{Uchiyama et al.}

\begin{document}

\title{\emph{Spitzer} IRAC Imaging of the Relativistic Jet 
from Superluminal Quasar PKS~0637$-$752}


\author{Yasunobu~Uchiyama,\altaffilmark{1} C.~Megan~Urry,\altaffilmark{1} 
Jeffrey~Van~Duyne,\altaffilmark{1} C.~C.~Cheung,\altaffilmark{2,3}
Rita~M.~Sambruna,\altaffilmark{4} Tadayuki~Takahashi,\altaffilmark{5,6}
Fabrizio~Tavecchio,\altaffilmark{7} and Laura~Maraschi\altaffilmark{7}  }

\altaffiltext{1}{Yale Center for Astronomy \& Astrophysics, 
Yale University, 260 Whitney Ave., New Haven, CT 06520-8121;
yasunobu.uchiyama@yale.edu,  meg.urry@yale.edu, vanduyne@astro.yale.edu}
\altaffiltext{2}{Jansky Postdoctoral Fellow; National Radio Astronomy
Observatory. Now hosted by Kavli Institute for Particle Astrophysics and 
Cosmology, Stanford University, Stanford, CA 94305}
\altaffiltext{3}{MIT Kavli Institute for Astrophysics \& Space Research,
77 Massachusetts Ave., Cambridge, MA 02139; ccheung@space.mit.edu}
\altaffiltext{4}{Department of Physics and Astronomy and School of 
Computational Sciences, George Mason University, 4400 University Drive,
Fairfax, VA 22030; rms@physics.gmu.edu}
\altaffiltext{5}{JAXA/ISAS, 3-1-1 Yoshinodai, Sagamihara, Kanagawa, 229-8510, Japan;
takahasi@astro.isas.jaxa.jp}
\altaffiltext{6}{Department of Physics, The University of Tokyo, 7-3-1 Hongo, Bunkyoku, 
Tokyo 113-0033, Japan}
\altaffiltext{7}{Osservatorio Astronomico di Brera, 
via Brera 28, Milano, I-20121, Italy; tavecchio@merate.mi.astro.it,
maraschi@brera.mi.astro.it}

\begin{abstract}
Emission from the relativistic jet located at hundreds of kpc from 
the core of the superluminal quasar 
PKS 0637$-$752 was detected at 3.6 and $5.8\ \mu\rm m$ 
with the Infrared Array Camera (IRAC) on the \emph{Spitzer Space Telescope}. 
The unprecedented sensitivity and arcsecond resolution 
of IRAC allows  us  to explore the mid-infrared 
 emission from kiloparsec-scale quasar jets for the first time.
The mid-infrared flux from the jet knots, when combined with  radio  and 
optical fluxes, confirms a synchrotron origin of 
 the radio-to-optical emission and constrains very well the high energy end
 of the nonthermal electron distribution. 
Assuming the X-rays are produced in the relativistically moving knots 
via inverse Compton 
scattering of cosmic microwave background (CMB) radiation,  
the infrared observation
puts constraints on the matter content of the quasar extended jet.
Specifically, pure $e^+e^-$ pair jet models are unfavorable 
based on the lack of an infrared bump associated with
  ``bulk Comptonization'' of CMB photons by an ultrarelativistic jet.  
\end{abstract}


\section{Introduction}

Since the discovery of an  X-ray jet of the quasar PKS 0637$-$752 
\citep{Cha00,Sch00}, 
a number of  X-ray jets extending to hundreds of kpc distances from 
the quasar nucleus have been unveiled by \emph{Chandra}
 \citep[see e.g.,][]{Sam04,Mar05}.
The X-ray emission mechanism, however, still remains unsettled  for 
most of the large-scale jets of powerful quasars.
Based on the spectral energy distribution (SED) from radio, optical, 
and X-ray bands, it has been argued that 
the X-ray intensity is too high to be explained by synchrotron 
 or synchrotron self-Compton radiation from a single population of electrons. 

A currently favored hypothesis for the strong X-ray emission is 
{\it relativistically-amplified} inverse Compton (IC)
scattering of cosmic microwave background (CMB),
in which the bulk flow of the jet is assumed to be relativistic,  
with a Lorentz factor $\Gamma \sim 10$ all the way to nearly Mpc distances,
and to be directed toward the observer at a small viewing angle 
of $\theta \sim \Gamma^{-1}$ \citep{Tav00,CGC01}. 
However, some difficulties in this model have been recognized, and 
alternative scenarios have been proposed  \citep{Aha02,DA02,SO02,AD04}.
Discussions of these models are hampered 
by poorly known physical conditions in quasar jets, as well as by limited observational 
windows  available  so far. 

In this Letter, we present \emph{Spitzer} IRAC imaging 
at wavelengths 3.6 and $5.8\ \mu\rm m$ of the jet of the superluminal quasar 
PKS 0637$-$752.
With the unprecedented sensitivity afforded by IRAC,
the observation aimed at measuring the mid-infrared part of the 
broadband nonthermal spectrum of the jet knots. Also we have searched for 
a possible infrared bump, which is expected if the jet is indeed 
 highly relativistic at hundreds kpc distances from the central engine, and
 contains cold (a Lorentz factor $\gamma \sim 1$) electrons.
The redshift of PKS 0637$-$752 is $z=0.651$ \citep{SBB76}, so we 
adopt a luminosity distance of $D_L = 1.26\times 10^{28}\ \rm cm$, 
for a $\Lambda$CDM cosmology with
$\Omega_{\rm m}=0.27$, 
$\Omega_{\Lambda}=0.73$, and 
$H_0=71\ \rm km\ s^{-1}\ Mpc^{-1}$.

\section{Observations and Results}
\label{observation}

\begin{figure*}  
\epsscale{0.8}
\plottwo{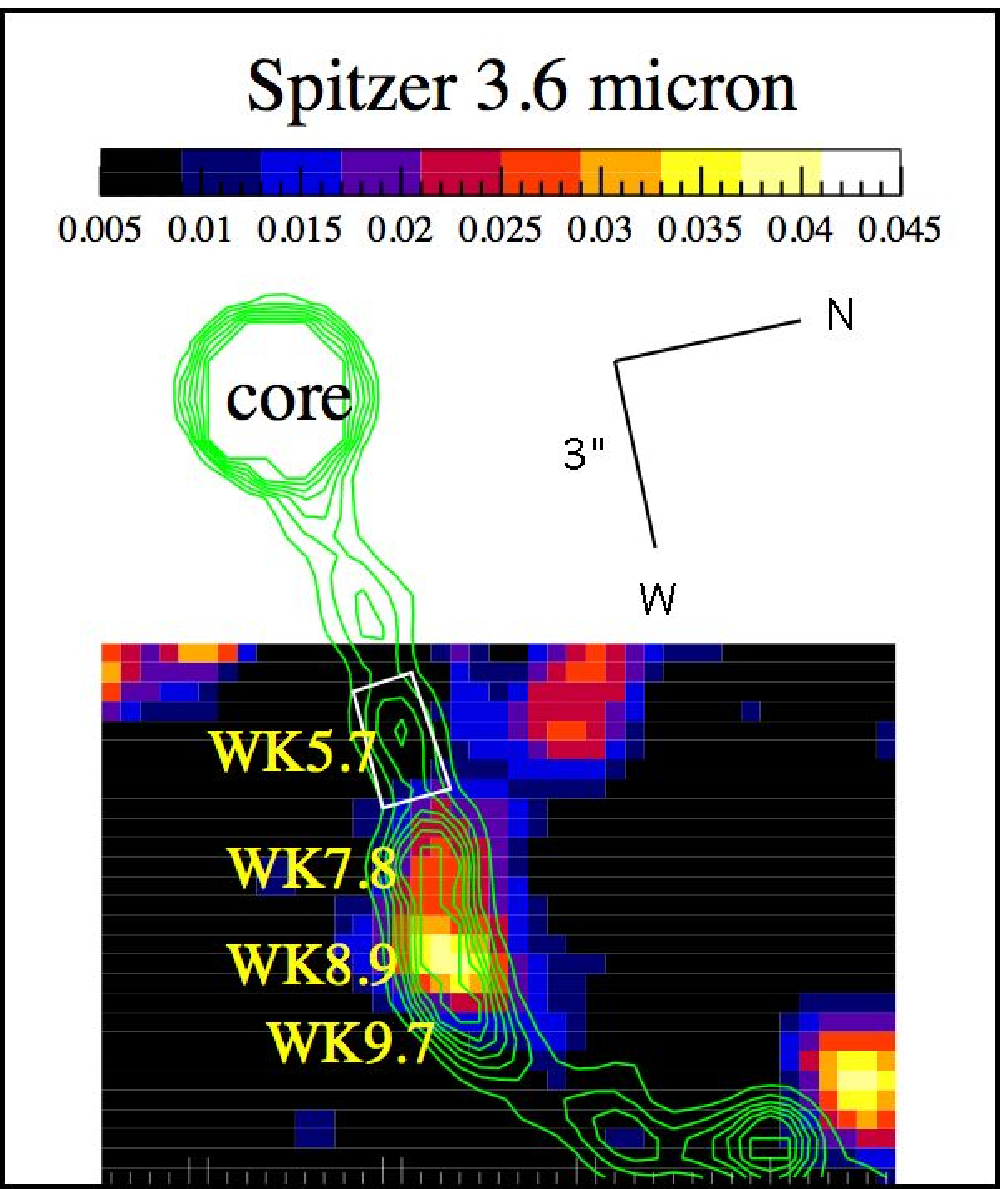}{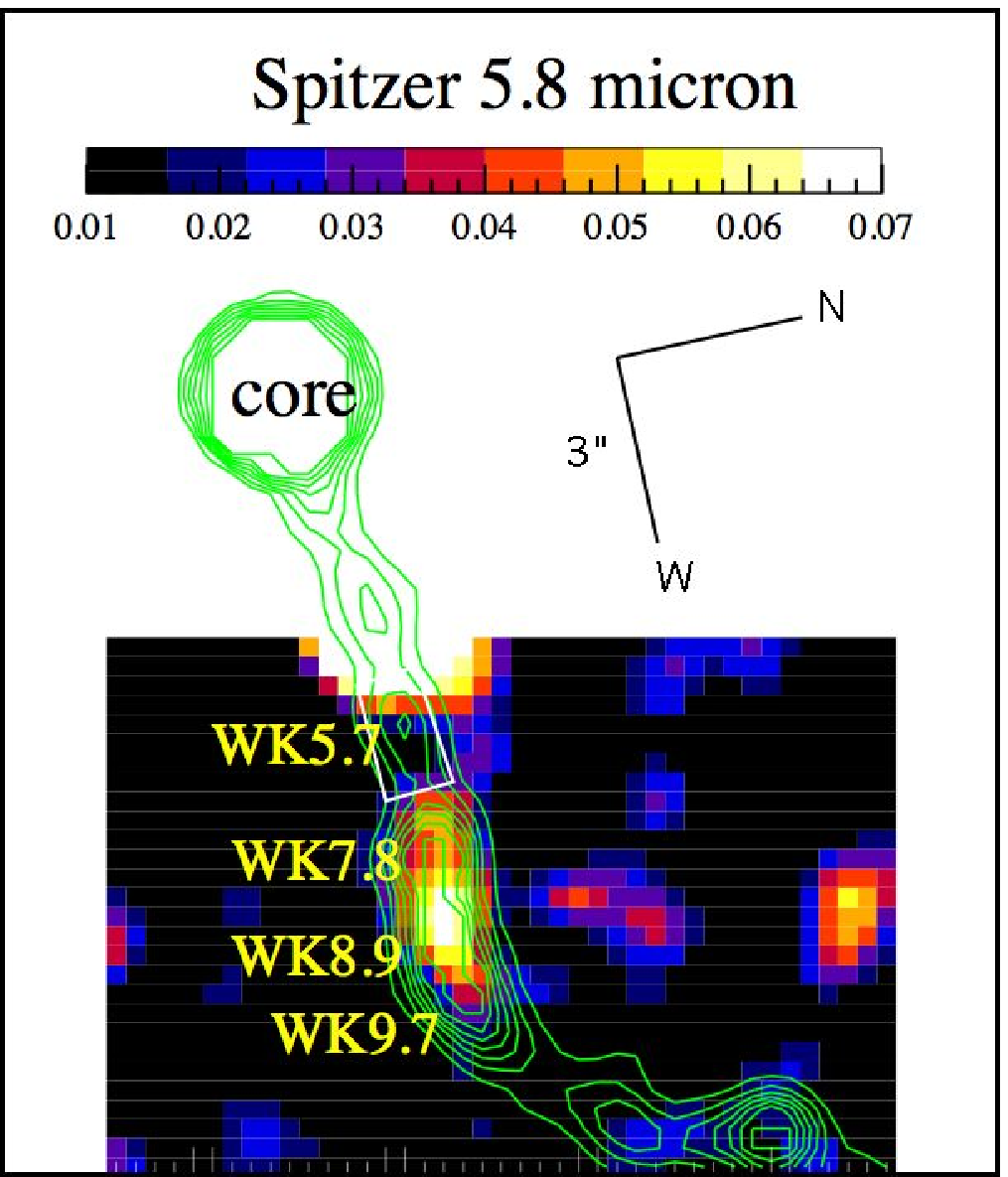}
\caption{IRAC images (\emph{false color}) of the PKS 0637$-$752 jet 
at $3.6\ \mu\rm m$ (left) and $5.8\ \mu\rm m$ (right) in units of 
$\rm MJy\ sr^{-1}$. An angular scale $1\arcsec$ corresponds to 6.9 kpc
projected size. 
Superimposed on the IRAC images are 
ATCA 8.6 GHz radio contours (using a circular restoring beam 
with $0\farcs88$ FWHM), starting at 
 $0.003\ \rm mJy\ beam^{-1}$ with an increment of  
 $0.003\ \rm mJy\ beam^{-1}$ \citep{Lov00}. 
The emission east of WK5.7 in the  $5.8\ \mu\rm m$ image 
is confused by residuals from wing cancellation. \label{fig:image} }
\end{figure*}

We have observed  PKS 0637$-$752 with \emph{Spitzer} IRAC 
on 2005 March 27 as part of our Cycle-1 General Observer program 
(\emph{Spitzer} Program ID 3586).
 IRAC is equipped with a four-channel camera,  
InSb arrays at 3.6 and $4.5\ \mu\rm m$ and 
Si:As arrays at  5.8 and $8.0\ \mu\rm m$, 
each with a $5\farcm2\times 5\farcm2$ field of view \citep{Faz04}.
Only the pair of 3.6 and $5.8\ \mu\rm m$ arrays,  observing 
the same sky simultaneously,  was chosen 
for the observation of PKS 0637$-$752, to obtain longer exposures in one pair 
of bandpasses as opposed to two pairs with truncated exposure time.
The 3.6/5.8 pair was chosen for a better spatial resolution and sensitivity.  
The pixel size in both arrays is $\simeq 1\farcs22$. The point spread 
functions are $1\farcs66$ and $1\farcs88$ (FWHM) for the 
3.6 and $5.8\ \mu\rm m$ bands, respectively \citep{Faz04}. 
A total of 50 frames per IRAC band, each with a 30-s frame time, were 
obtained.
The 50 images of the Basic Calibrated Data 
processed in the \emph{Spitzer} Science Center 
(SSC) were combined into a projected image using the SSC 
software {\sf mopex}.

Our aims are to explore possible infrared emission of the jet 
close ($<12\arcsec$)  to the quasar core. 
For this purpose, 
 the wings of the bright core must be subtracted carefully. 
We made use of the point response function (PRF) image of a bright star
provided by the SSC  to cancel out the PRF wings.
We also employed 
three bright field stars in our  $3.6\ \mu\rm m$ image as PRF templates, 
and found that the results from the SSC template are 
 consistent with those from the field star templates. 
 In our $5.8\ \mu\rm m$ image, on the other hand, 
there are no suitable stars for such comparisons. 
In what follows, we present both the 3.6 and $5.8\ \mu\rm m$ results 
making use of the SSC template alone. 
The infrared fluxes from the core  of  PKS 0637$-$752  were measured as 
5.0 and 9.6 mJy in the $3.6$ and $5.8\ \mu\rm m$ bands, respectively.
These values are below the saturation limits for a 30-s frame.

Figure \ref{fig:image} shows the IRAC images of 
 PKS 0637$-$752 after subtraction of the PRF wings of the 
 core.
The pixel size of the projected images of both channels was set to be 1/4 of 
 $1\farcs22$.
Regions close to the quasar core are blanked out due to 
residuals in the wing-subtraction.
Mid-infrared counterparts of the main optical knots, WK7.8 and WK8.9 \citep{Sch00}, which are also the main X-ray features, 
are clearly visible in both channels,  located $\sim 8\arcsec$ west of the core.
The two optical knots are marginally separated in the infrared images. 
In the  $3.6\ \mu\rm m$ image, WK8.9 appears to be brighter than  WK7.8, 
as in the optical. In the X-ray, WK8.9 is brighter than WK7.8 but 
the fluxes are more similar \citep{Cha00}. 
No significant infrared emission can be seen at WK5.7, 
WK9.7 ($\sim 3$ times fainter than WK8.9 in the optical), 
or farther components after a bending point at $\sim 10\arcsec$ west of the quasar.

Photometry was performed  with $5\arcsec$ diameter apertures enclosing 
both  WK7.8 and WK8.9, which finds $f_{3.6} = 6.5\pm0.5\ \mu\rm Jy$ and 
$f_{5.8} = 11\pm2\ \mu\rm Jy$ for 3.6 and  $5.8\ \mu\rm m$, respectively.
The $1\sigma$ errors include the uncertainties associated with PRF removal,  
adopting 10\% of the quasar's wing intensity at the location being considered. 
Note that we measure the sum of infrared flux from the two optical knots.
The infrared-to-optical slope\footnote{We define  
$\alpha_{\rm io} \equiv - \log (f_{\rm ir}/f_{\rm op})/\log (\nu_{\rm ir}/\nu_{\rm op})$,  
with $\nu_{\rm ir}=0.83\times 10^{14}\ \rm Hz$ (for $3.6\ \mu\rm m$)
and $\nu_{\rm op}=4.3\times 10^{14}\ \rm Hz$.
We take $f_{\rm ir}=6.5\ \mu\rm Jy$ and 
$f_{\rm op}=0.48\ \mu\rm Jy$ \citep{Sch00}.}
is $\alpha_{\rm io} \simeq1.6$, significantly steeper than the spectral index 
at either radio or 
X-ray wavelengths, i.e., $\alpha_{\rm r} =0.81 \pm 0.01$ and 
$\alpha_{\rm X} =0.85 \pm 0.08$ \citep{Cha00}.

Although the separation of the two knots is only $\simeq 1\arcsec$,
we can derive a rough flux ratio 
between WK7.8 and WK8.9 at the  $3.6\ \mu\rm m$ band,
 thanks to the high significance and better PRF in this band. 
The flux from WK8.9 is higher by a factor of $1.5\pm0.6$ than 
WK7.8, in  agreement with a factor of $\simeq 1.4$ in the optical. 

We can place upper limits on the IR fluxes from a ``quiet" region preceding 
the nonthermal condensations WK7.8--8.9. 
For a white box in Fig.~\ref{fig:image} that encloses WK5.7, 
$f_{3.6}\!<\! 2\ \mu\rm Jy$ and $f_{5.8} \!<\! 5\ \mu\rm Jy$ (at a $2\sigma$ level) 
are obtained. 
The flux uncertainty is computed as $2\sqrt{N} \sigma$, where 
$N$ is  the ``noise pixels"  taken from \citet{Faz04} and $\sigma$ presents 
the background noise in which PRF-removal uncertainty is also formally included. 
We use these flux limits later in \S\ref{BC}.

\section{Discussion} 
\label{discuss}

\subsection{SED of Knot WK7.8}
\label{SED}

\begin{figure}
\epsscale{1.0}
\plotone{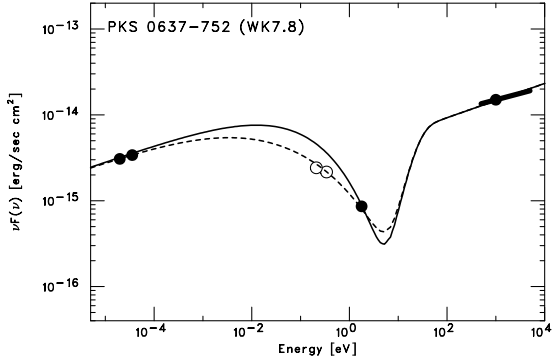}
\caption{Broadband SED for knot WK7.8. IRAC 3.6 and 
$5.8\ \mu\rm m$ fluxes, 2.7 and $4.6\ \mu\rm Jy$, respectively, are shown 
as open circles. 
The radio flux at 8.6 GHz, 44 mJy, is extracted from the ATCA map 
centered on WK7.8 with a radius $0\farcs8$, while the 4.8 GHz flux is set 
to satisfy $\alpha_{\rm r}=0.8$. 
The \emph{HST} optical ($0.2\ \mu\rm Jy$ at 697 nm) and \emph{Chandra} 
 X-ray data (6.2 nJy at 1 keV) are taken from the list of \citet{KS05}.
The solid and dashed lines represent the synchrotron$+$IC/CMB modeling 
(see text).  \label{fig:SED}}
\end{figure}

The mid-infrared fluxes that we have measured fill central points in the SED 
of the jet knots and, therefore, set an important constraint on models 
of the broadband emission.
In Fig.~\ref{fig:SED}, we show the SED  for knot WK7.8, 
the first bright  knot which has been frequently modeled in the literature.
The knot is unresolved both in the radio image (ATCA) and the \emph{HST} image
\citep{Sch00}. Also the corresponding knot is visible in the 
deconvolved \emph{Chandra} X-ray image \citep{Cha00}. The emissions  in these 
bands  most likely emerge  from the same physical volume. 
(In contrast,  knot WK8.9 appears elongated along the jet in the optical, and so
 may involve multiple knots.) 
Setting a flux ratio in the infrared between WK7.8 and WK8.9 
to be 1.4 based on the optical (see above), we associate flux densities of 
$2.7\ \mu\rm Jy$ (at $3.6\ \mu\rm m$) and $4.6\ \mu\rm Jy$ (at $5.8\ \mu\rm m$)
with knot WK7.8.

The radio-to-optical spectrum is now arguably  attributable to synchrotron radiation, 
because extrapolation of the radio spectrum can smoothly connect with 
the infrared and optical fluxes assuming normal steepening due to radiative cooling.
Specifically, we first assume  that 
the energy distribution of the electrons (in the jet co-moving frame) 
follows  
$N(\gamma) = K \gamma^{-s} \exp (-\gamma/\gamma_{\rm max})$ 
for $\gamma \!>\! \gamma_{\rm min}$,
where $K$ is the normalization,  
$\gamma$ denotes the Lorentz factor of relativistic electrons, and 
 $s=2.6$ to match the radio slope $\alpha \simeq 0.8$ ($s=2\alpha+1$). 
To reproduce the optical flux, we obtain  
$\gamma_{\rm max} \simeq 3.6\times 10^5 \,(\delta_{10}B_{-5})^{-1/2}$, 
where $\delta_{10}=\delta/10$ and 
$B_{-5}=B/(10^{-5}\rm G)$ are the characteristic 
Doppler factor\footnote{
The Doppler factor is defined as $\delta \equiv [\Gamma(1-\beta\cos\theta)]^{-1}$ 
where $\beta c$ the velocity of the jet, 
 $\Gamma = (1-\beta^2)^{-1/2}$ the bulk Lorentz factor of the jet, 
 and $\theta$ is the observing angle with respect to the jet direction.} and 
co-moving magnetic field strength (the solid line in Fig.~\ref{fig:SED}).
The low energy cutoff $\gamma_{\rm min}$ is arbitrary as long as 
$\gamma_{\rm min} \!<\!  2\times 10^3 \,(\delta_{10}B_{-5})^{-1/2}$.

This simple model for the synchrotron component overpredicts the mid-infrared 
flux by a factor of $\sim 2$.
To account for this discrepancy, 
we modified the form of the electron distribution to 
$N(\gamma) = K \gamma^{-s} (1+\gamma/\gamma_{\rm br})^{-1}
\exp (-\gamma/\gamma_{\rm max})$ with $s=2.6$, 
which is a smooth broken power law with an exponential cutoff, taking 
account of possible synchrotron and IC (in the Thomson regime) cooling.  
Fitting the SED with this model, we obtain 
$\gamma_{\rm br} \simeq 1.4\times 10^5 \,(\delta_{10}B_{-5})^{-1/2}$ and 
$\gamma_{\rm max} \simeq 1.1\times 10^6 \,(\delta_{10}B_{-5})^{-1/2}$
(the dashed line in Fig.~\ref{fig:SED}).

Here we also model the X-ray spectrum, for illustrative purpose, by 
the IC/CMB model \citep{Tav00}, 
namely by the model invoking relativistically-amplified IC scattering on CMB radiation, 
with the Doppler factor of $\delta =8.4$ [corresponding to case (b) in \S\ref{BC}]
and $\gamma_{\rm min} =20$. In order to produce the X-ray flux and 
at the same time to suppress the optical flux by the IC process, 
we obtain a tight constraint, $10 \la \gamma_{\rm min} \la 40$.
Note that the IC/CMB model for the X-rays generally requires large power carried by 
nonthermal relativistic electrons, 
$L_{\rm knot} \simeq \pi r^2 \Gamma^2 \beta c\, u_e$,  
where $r=1\ \rm kpc$ is the adopted radius of the knot and 
$u_e$ is the energy density of nonthermal electrons in the jet frame, 
which can be written as 
$u_e=\langle \gamma \rangle m_ec^2\,n_e = 
[(p\!-\!1)/(p\!-\!2)]\,\gamma_{\rm min} m_ec^2\, n_e$ with 
the number density of nonthermal electrons $n_e$.
For example, one needs 
$L_{\rm knot} \sim 9.2 \times 10^{46}\ \rm ergs\ s^{-1}$ in the case of 
the parameter set (b) in \S\ref{BC} 
($\delta =8.4$, $\Gamma =12$, and $\gamma_{\rm min} =20$).

The detected mid-infrared fluxes well constrain 
the highest energy part of the electron distribution, 
 roughly $\gamma \sim 10^{5\mbox{--}6}$. 
In addition, as we show below, 
the \emph{non-detection} of infrared light from the jet preceding WK7.8 
strongly  constrains the 
lowest energy ($\gamma \sim 1$) electrons, which have yet to be accelerated.  

\subsection{Bulk Compton Radiation}
\label{BC}

The relativistically-amplified IC/CMB scenario raises the interesting possibility that 
an infrared bump may appear as a result of 
  ``bulk Comptonization'' (hereafter BC), namely 
Comptonization of the CMB photons by  ``cold" 
($\gamma \sim 1$) electrons or positrons in the 
ultrarelativistic  jet stream \citep{Sik97,Geo05}. 
The BC bump is expected to have a blackbody-like spectrum peaking 
at the energy $\varepsilon_{\rm BC} \simeq 2\delta\Gamma \varepsilon_{\rm CMB}$, 
independent of redshift, where $\varepsilon_{\rm CMB}$ is the mean energy of the CMB 
radiation at $z=0$.
For $\Gamma = \delta = 10$, the bump appears 
at the infrared wavelength $\lambda_{\rm BC} \sim 10\  \mu \rm m$.
The isotropic BC luminosity from relativistic jet containing cold electrons,  
assuming a cylindrical geometry of radius $r$ and length $l_{\rm cold}$, 
is given by \citet{Geo05}:
\begin{equation}
L_{\rm BC}  \simeq \frac{4}{3} \frac{\sigma_{\rm T}}{m_ec^2} 
l_{\rm cold} U_{\rm CMB} \beta \delta^{3}\ L_{\rm cold}, 
\end{equation}
where $\sigma_{\rm T}$ is the Thomson cross section, 
$U_{\rm CMB} =4.2\times 10^{-13} (1+z)^4\ \rm ergs\ cm^{-3}$, 
and  $L_{\rm cold}$ is the jet power carried by cold electrons 
that upscatter CMB radiation into the infrared, 
defined as $L_{\rm cold} = \pi r^2 \Gamma^2 \beta c
n_{\rm cold} m_e c^2$ with the number density of cold electrons in the jet frame,
$n_{\rm cold}$.

\begin{figure}
\epsscale{1.0}
\plotone{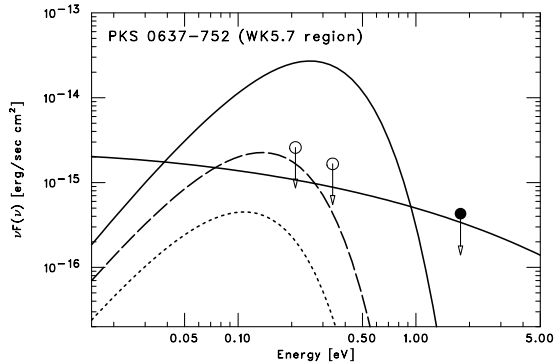}
\caption{Bulk Comptonization spectra expected from the WK5.7 region 
($2\arcsec$ long)  assuming $L_{\rm cold} = 10^{46}\ \rm ergs\ s^{-1}$ 
and 
(a) 
$\Gamma = 12$, $\theta = 3.5\degr$ ({\it solid curve}), 
(b) 
$\Gamma = 12$, $\theta = 6.5\degr$ ({\it dashed curve}), 
(c)
$\Gamma = 15$, $\theta = 8.3\degr$ ({\it dotted curve}). 
Also shown is a synchrotron spectrum (solid horizontal line) 
 extrapolated  from the 
radio flux at 8.6 GHz with the spectral form of the dotted line in Fig.~\ref{fig:SED}. 
The upper limits on IRAC 3.6 and 
$5.8\ \mu\rm m$ fluxes are shown as open circles. The optical 
upper limit is obtained from the 696.9 nm \emph{HST} image \citep{Sch00}.
\label{fig:BC}}
\end{figure}

Along a ``quiet''  stream between the core and the first bright knot WK7.8, 
it is expected that 
unseen cold particles are propagating and generating BC radiation. 
In Fig.~\ref{fig:BC}, we present BC spectra 
calculated for the quiet portion of the jet, 
assuming for illustrative purpose that 
$L_{\rm cold} = 10^{46}\ \rm ergs\ s^{-1}$.
The length of this region is set to be 
$l_{\rm cold} = (13.8/\sin\theta)\ \rm kpc$ appropriate for the white box 
 in Fig.~\ref{fig:image},
  where the flux upper limits were derived in \S\ref{observation}. 
We consider the three characteristic cases as follows:
(a) 
$\Gamma = 12$, $\theta = 3.5\degr$,
(b) 
$\Gamma = 12$, $\theta = 6.5\degr$,
(c)
$\Gamma = 15$, $\theta = 8.3\degr$.
These parameters are chosen for the consistency with 
the VSOP and VLBI observations, which   found superluminal motions 
on milliarcsecond scale ($\sim 100$ pc distances from the core) 
with a mean apparent speed of $\beta_{\rm app}c = 11.4\pm0.6\, c$, indicating 
$\Gamma >11.4$ \citep{Lov00}.
The viewing angles are set by the well-known relation \citep[see][]{UP95} 
$\beta_{\rm app} = \beta \sin\theta/(1\!-\!\beta \cos\theta)$ with 
 $\beta_{\rm app} = 11.4$,  which yields 
two solutions for given $\Gamma$. 
For $\Gamma=15$, the smaller angle solution $\theta = 1.8\degr$ 
 requires an unlikely long jet,  
exceeding a de-projected total length of 2.2 Mpc, and therefore is omitted. 
From Fig.~\ref{fig:BC}, one can constrain $L_{\rm cold}$
by comparing the calculated BC spectra to the flux upper limits 
obtained from the IRAC data. The results are  summarized in Table \ref{tbl-1}.

In Table \ref{tbl-1},  the jet power in nonthermal electrons, 
$L_{\rm knot}$, is given 
based on the IC/CMB modeling of  knot WK7.8  (with $\gamma_{\rm min}=20$), 
together with the lower limits on 
$\xi \equiv L_{\rm knot}/L_{\rm cold} = \langle \gamma \rangle n_e/n_{\rm cold}$.
The ratio of the energy densities of nonthermal electrons $u_e$ and magnetic 
fields $u_B$ is also shown.
It appears that case (c) has uncomfortably large $L_{\rm knot}$ and $u_e/u_B$ 
because of small $\delta$ \citep[see also][]{DA04}.

\begin{table}
\begin{center}
\caption{Jet powers in cold and nonthermal electrons.\label{tbl-1}}
\begin{tabular}{ccccccccc}
\tableline\tableline
Case & $\Gamma$ & $\theta$ & $\delta$ & $L_{\rm cold}$ & $L_{\rm knot}$ &  
$\xi$  & $n_{\rm cold}/n_e$ & $u_e/u_B$ \\ 
 & & (deg)  & & $(\rm ergs\ s^{-1})$ &$(\rm ergs\ s^{-1})$ &
 &  & \\
\tableline
(a) & 12 & $3.5$ & 15.6 & $<\! 1.0\times10^{45}$ & $2.8\times10^{45}$ 
& $>\! 2.8$ & $<\! 20$ &$0.69$ \\
(b) & 12 & $6.5$ & 8.4 & $<\! 1.6\times10^{46}$ &  $9.2\times10^{46}$ 
& $>\! 5.8$ & $<\! 10$ & $92$ \\
(c) & 15 & $8.3$ & 5.3 &  $<\! 1.4\times10^{47}$ &  $1.9\times10^{48}$ 
& $>\! 14$ & $<\! 4$ & $2.9\times 10^3$ \\
\tableline
\end{tabular}
\end{center}
\end{table}

The limits on $L_{\rm cold}$ impose meaningful constraints on the jet models.
If the jet is dynamically dominated by $e^+e^-$ pairs, one can assume that  
the energy flux of nonthermal electrons at the knot 
is stored in cold $e^+e^-$ pairs before the first knot, therefore
 $\xi \leq 1$ (i.e.\ $L_{\rm cold} \geq L_{\rm knot}$). 
However, we obtain $\xi > 1$ for all cases (see Table \ref{tbl-1}).
Our estimates of $\xi$ cannot be noticeably  reduced 
by increasing $\gamma_{\rm min}$ 
(through the relation $L_{\rm knot} \propto \gamma_{\rm min}^{-0.6} $)
 as $\gamma_{\rm min} \la 40\, \delta_{10}^{-1}$ is set by the X-ray spectrum.
Also, changing $\theta$ in a plausible range $3\degr \!\leq  \theta  \leq  9\degr$
both for $\Gamma=12$ and 15 
does not affect our conclusion in this regard. 
Therefore, as long as  the IC/CMB scenario is correct, 
our result argues against pure $e^+e^-$ jet models.

Interestingly, 
{\it on  sub-pc scales}  in radio loud quasars, 
a similar conclusion has been reached by \citet{SM00} that 
jet composition with pure $e^+e^-$ pairs can be excluded 
by the absence of BC radiation at X-ray energies 
due to Comptonization of UV photons. 
The present work is the first time that jet composition on scales of 
many kiloparsecs has been constrained in this way.

The number density of cold electrons relative to that of nonthermal 
electrons can be constrained as $n_{\rm cold}/n_e \! <\! \mbox{4--20}$ 
(see Table \ref{tbl-1}),  through 
the relation $\xi = \langle \gamma \rangle n_e/n_{\rm cold}$ 
and $\langle \gamma \rangle \simeq 2.7 \gamma_{\rm min} \sim 50$.
We note that these limits are difficult to explain within the standard 
picture of particle acceleration, namely diffusive shock acceleration, 
in which 
it is generally expected that 
only a small fraction of the cold electrons can be accelerated to nonthermal 
energies, and  consequently  $n_{\rm cold}/n_e \!\gg\! 1$. 
One may need to invoke alternative models regarding 
the X-ray emission mechanism (namely IC/CMB) and/or the acceleration 
mechanism. 

A possible caveat is that 
the above discussion relies on the simplified picture that 
the  power transported by the jet is constant. If the ejection of power at 
the base of the jet is discontinuous, the emission knots may represent 
``power peaks''  and the average jet power could be smaller than 
the value of $L_{\rm knot}$ we calculated. 
It should also be noted that  in our calculations of the BC spectra, we have assumed monoenergetic 
cold electrons of $\gamma =1$. 
If instead a certain fraction of pairs is ``hot" 
 with $\gamma$ of a few,  the BC spectrum can be  different from 
our calculation, extending to optical 
energies. Even so, because of the lower limit on the optical flux, 
the limits on  $L_{\rm cold}$ would be similar 
to, or tighter than,  what we obtained above.


\acknowledgments

We are grateful to Jim Lovell for providing us the ATCA 8.6 GHz radio image.
We would like to thank the anonymous referee for very helpful comments. 
This work was supported in part by NASA grant NAG5-12873.
This work is based on observations made with the Spitzer Space Telescope,
which is operated by the Jet Propulsion Laboratory, California Institute of 
Technology under NASA contract 1407. 


\begin{thebibliography}{}

\bibitem[Aharonian(2002)]{Aha02} Aharonian, F. A. 
2002, \mnras, 332, 215 
\bibitem[Atoyan \& Dermer(2004)]{AD04} Atoyan, A. M., \& Dermer, C. D.
2004, \apj, 613, 151
\bibitem[Celotti, Ghisellini, \& Chiaberge(2001)]{CGC01} Celotti, A.,
Ghisellini, G., \& Chiaberge, M.
2001, \mnras, 321, L1 
\bibitem[Chartas et al.(2000)]{Cha00} Chartas, G., et al.\ 
2000, \apj, 542, 655
\bibitem[Dermer \& Atoyan(2002)]{DA02} Dermer, C. D., \& Atoyan, A. M.
2002, \apjl, 568, L81
\bibitem[Dermer \& Atoyan(2004)]{DA04} Dermer, C. D., \& Atoyan, A. M.
2004, \apjl, 611, L9
\bibitem[Fazio et al.(2004)]{Faz04} Fazio, G., et al.\
2004, \apjs, 154, 10
\bibitem[Georganopoulos et al.(2005)]{Geo05} Georganopoulos, M.,
Kazanas, D., Perlman, E., \& Stecker, F. W.
2005, \apj, 625, 656
\bibitem[Kataoka \& Stawarz(2005)]{KS05} Kataoka, J., \& Stawarz, \L.
2005, \apj, 622, 797
\bibitem[Lovell et al.(2000)]{Lov00} Lovell, J. E. J., et al.\
2000, in Astrophysical Phenomena Revealed by Space VLBI, ed.\ H. Hirabayashi,
P. G. Edwards, \& D. W. Murphy (Sagamihara: ISAS), 215
\bibitem[Marshall et al.(2005)]{Mar05} Marshall, H. L., et al.\
2005, \apjs, 156, 13
\bibitem[Sambruna et al.(2004)]{Sam04} Sambruna, R. M., 
Gambill, J. K., Maraschi, L., et al.\
2004, \apj, 608, 698
\bibitem[Savage, Browne, \& Bolton(1976)]{SBB76} Savage, A., 
Browne, I.~W.~A., \& Bolton, J.~G.\ 1976, \mnras, 177, 77P
\bibitem[Schwartz et al.(2000)]{Sch00} Schwartz, D. A., et al.\
2000, \apjl, 540, L69
\bibitem[Sikora \& Madejski(2000)]{SM00} Sikora, M., \& Madejski, G.
2000, \apj, 534, 109
\bibitem[Sikora et al.(1997)]{Sik97} Sikora, M., Madejski, G., 
Moderski, R., \& Poutanen, J.
1997, \apj, 484, 108
\bibitem[Stawarz \& Ostrowski(2002)]{SO02} Stawarz, \L., \& Ostrowski, M.
2002, \apj, 578, 763
\bibitem[Tavecchio et al.(2000)]{Tav00} Tavecchio, F., 
Maraschi, L., Sambruna, R. M., Urry, C. M. 
2000, \apjl, 544, L23
\bibitem[Urry \& Padovani(1995)]{UP95} Urry, C. M., \& Padovani, P.
1995, \pasp, 107, 803


\end{thebibliography}
\end{document}